\documentclass{aastex631}
\usepackage{amsmath}
\usepackage{mathrsfs}
\begin{document}

\title{Updated estimates of post-merger gravitational wave energy released in GW170817 and the maximum neutron star mass}
\author{Yong Chen}
\affiliation{Key Laboratory of Dark Matter and Space Astronomy, Purple Mountain Observatory, Chinese Academy of Sciences, Nanjing 210033, China}
\affiliation{School of Astronomy and Space Science, University of Science and Technology of China, Hefei, Anhui 230026, China}
\author{Bo Gao}
\affiliation{Key Laboratory of Dark Matter and Space Astronomy, Purple Mountain Observatory, Chinese Academy of Sciences, Nanjing 210033, China}
\author[0000-0002-7505-7795]{Yong-Jia Huang}
\affiliation{Key Laboratory of Dark Matter and Space Astronomy, Purple Mountain Observatory, Chinese Academy of Sciences, Nanjing 210033, China}
\author[0000-0001-9120-7733]{Shao-Peng Tang}
\affiliation{Key Laboratory of Dark Matter and Space Astronomy, Purple Mountain Observatory, Chinese Academy of Sciences, Nanjing 210033, China}
\author[0000-0002-8966-6911]{Yi-Zhong Fan}
\affiliation{Key Laboratory of Dark Matter and Space Astronomy, Purple Mountain Observatory, Chinese Academy of Sciences, Nanjing 210033, China}
\affiliation{School of Astronomy and Space Science, University of Science and Technology of China, Hefei, Anhui 230026, China}

\begin{abstract}
In this work, benefiting from the increased sample of neutron stars with measured masses and radii as well as the incorporation of the chiral effective field theory and perturbative QCD constraints, the tidal parameter $\kappa_2^T$ is constrained to be $78^{+17}_{-11}$ (68.3\% credible interval, mainly adopted in this work unless mentioned specifically) for the binary neutron stars involved in GW170817. Such a $\kappa_2^T$ is in favor of strong gravitational wave radiation in the post-merger phase and the corresponding energy is estimated to be $E_{\rm GW,p} \simeq 0.051^{+0.022}_{-0.017}\,M_\odot c^2$. Assuming the remnant from binary neutron star merger is a supramassive neutron star, as suggested by the modeling of the electromagnetic counterparts of GW170817, we examine the maximum mass of nonrotating neutron stars ($M_{\rm TOV}$) while accounting for the uncertainty in the remnant's lifetime ($t_{\rm c}$). Our results show that $M_{\rm TOV}$ varies from $2.09^{+0.11}_{-0.09}\,M_\odot$ for $t_{\rm c}=0.1$~s to $2.18^{+0.10}_{-0.09}\,M_\odot$ for $t_{\rm c}=1$~s.
This result is consistent with that independently inferred from the re-construction of the equation of state of neutron star matter, i.e., $M_{\rm TOV,exc}=2.16^{+0.10}_{-0.07}M_\odot$, particularly if the very massive neutron stars with masses measured indirectly have been removed in constructing the prior distribution of $M_{\rm TOV}$. The consistency of the maximum mass of nonrotating neutron stars found in different approaches suggests a reasonable understanding of this key parameter for dense-matter physics.
\end{abstract}

\section{Introduction}
\label{sec:intro}
Neutron star (NS) mergers are sensitive probes of dense matter physics through both their gravitational waves (GWs) and remnant properties \citep{Baiotti:2019sew,Shibata:2019wef,Radice:2020ddv,Dietrich:2020eud}. In particular, the maximum mass of a cold, non-rotating NS ($M_\mathrm{TOV}$) is fixed by the Tolman–Oppenheimer–Volkoff equations for a given equation of state (EoS) \citep{Oppenheimer:1939ne}, so a measurement of $M_\mathrm{TOV}$ would strongly constrain the EoSs proposed currently \citep{Lattimer:2012nd}. The value of $M_{\rm TOV}$ also sets the threshold between prompt collapse and a (possibly long-lived) NS remnant in a binary neutron star (BNS) merger, and therefore impacts the observable signals, e.g., GWs, gamma-ray bursts, and kilonovae \citep{Baiotti:2016qnr,Bernuzzi:2020tgt}. Theoretically, causality arguments allow $M_{\rm TOV}\lesssim3.2\,M_\odot$ \citep{Rhoades:1974fn}, and nuclear models compatible with current astrophysical and nuclear-physics constraints suggest somewhat lower limits (e.g., $\lesssim2.5\,M_\odot$, \citealt{Oertel:2016bki}). Observationally, several radio pulsars exceed $2\,M_\odot$, placing a robust lower bound on $M_{\rm TOV}$ \citep[e.g., PSR J0740+6620,][]{Fonseca:2021wxt}. Likewise, the total energy emitted in post-merger GWs ($E_{\rm GW,p}$) can be a few percent of the binary's mass \citep{Bernuzzi:2015opx, Zappa:2017xba}, carrying information on the remnant’s dynamics and EoS \citep{Takami:2014zpa,Huang:2022mqp}. Constraining $E_{\rm GW,p}$ is therefore crucial: for example, if more energy is radiated away in GWs, then less angular momentum remains in the remnant to support it against collapse, which in turn affects the inference of $M_{\rm TOV}$ \citep{Fan:2013cra,Shibata:2019ctb,Shao:2019ioq}.

Although post-merger GW emission was not detected in GW170817 \citep{LIGOScientific:2017fdd}, theoretical models suggest that substantial energy could be radiated in the post-merger phase \citep{Zappa:2017xba, Torres-Rivas:2018svp}. Indeed, using a combination of the GW data of GW170817 and the NICER mass--radius measurement of PSR J0030+0451, \citet{Fan:2020hwe} constrained the EoS and found that the properties of the compact objects involved in GW170817 are likely in favor of a strong post-merger GW emission. Efficient post-merger emission would yield a relatively high $M_{\rm TOV}$, which is helpful in loosening the potential tension with the (indirectly-measured) high masses of some massive pulsars.

Although no additional reliable BNS merger events have been detected in recent years, the NICER collaboration has made significant progress in measuring the radii of the NSs. Following PSR J0030+0451, NICER measured the massive pulsar PSR J0740+6620 \citep{Riley:2021pdl, Miller:2021qha}, whose independently determined mass provides a valuable high-mass benchmark \citep{Fonseca:2021wxt}. Updated analyses using several years of NICER data combined with XMM-Newton observations found that the inferred radius of PSR J0740+6620 remains consistent with earlier results \citep{Dittmann:2024mbo, Salmi:2024aum}. This is however not the case for PSR J0030+0451, for which the subsequent re-analysis, incorporating improved instrument calibration and expanded data sets, showed that while the compactness of PSR J0030+0451 is relatively well constrained, its radius estimate depends sensitively on assumptions about surface hot-spot geometry \citep{Vinciguerra:2023qxq}. Fortunately, \citet{Luo:2024lbz} (see also \citealt{Biswas:2025ivu}) found that the ST+PDT hot-spot configuration (in which one hot spot is modeled as a single-temperature circular region while the other consists of two overlapping components with different temperatures) is statistically preferred. More recently, NICER has released $M$–$R$ measurements for additional pulsars, including for example PSR J0437–4715 with precise mass determinations \citep{Choudhury:2024xbk}. Such a growing set of well-observed pulsars, including PSR J0030+0451, J0740+6620, and PSR J0437–4715, collectively strengthens the empirical basis for EoS inference by reducing biases associated with any single source.

Incorporating these new measurements would therefore enable a more precise EoS reconstruction, which in turn allows improved inference of the post-merger GW energy $E_{\mathrm{GW,p}}$ and the maximum nonrotating NS mass $M_{\mathrm{TOV}}$, when combined with the multi-messenger observations of GW170817 \citep{LIGOScientific:2017vwq,LIGOScientific:2017ync} and quasi-universal relations (QURs, which are EoS-insensitive). In particular, empirical fits relate the binary tidal parameter $\kappa_2^T$ to the fraction of energy radiated during the post-merger phase \citep{Zappa:2017xba}, and universal relations connect the critical collapse mass to $M_{\rm TOV}$ \citep{Breu:2016ufb,Shao:2019ioq}. Updating these constraints using the latest observational data therefore constitutes the primary motivation of this work. Since the lifetime of the remnant of GW170817 is still under debate, in this work we also treat the collapse time as a free parameter and quantify its impact on the inferred $M_{\rm TOV}$.

\section{Updated estimate of the post-merger GW radiation of GW170817} \label{sec:second}
It has been shown that the leading-order effective-one-body (EOB) tidal coupling parameter $\kappa_2^T$ \citep{Damour:2012yf} provides an effective characterization of tidal interactions during the late inspiral and merger phases of BNS coalescences \citep{Bernuzzi:2015rla}. Let $\Lambda$ denote the dimensionless tidal deformability of each component NS, which is related to the quadrupolar Love number $k_2$ through $\Lambda = 2/3\,\mathscr{C}^{-5} k_2$, where $\mathscr{C}$ is the stellar compactness. The parameter $\kappa_2^T$ is then defined as $\kappa_2^T = 3\,(M_1^4 M_2 \Lambda_1 + M_2^4 M_1 \Lambda_2)/(M_1 + M_2)^5$, which encodes information about the NS EoSs and the component masses. It naturally correlates with both the characteristic GW frequency during merger \citep{Bernuzzi:2014kca,Bernuzzi:2015rla} and the amount of GW energy radiated in the early post-merger phase \citep{Bernuzzi:2015opx}. The reduced post-merger GW energy is $e_{\rm GW,p} \equiv E_{\rm GW,p}/(M_{\rm tot} \nu c^2)$, where $\nu = M_1 M_2/M_{\rm tot}^2$ is the symmetric mass ratio and $M_{\rm tot}=M_1+M_2$ is the total gravitational mass. Through a large set of numerical relativity simulations, \citet{Zappa:2017xba} provided the following best-fit relation between $e_{\rm GW,p}$ and $\kappa_2^T$:
\begin{equation}
e_{\rm GW,p}( \kappa^T_2 ) =
        \begin{cases}
            0.02  & \kappa^T_2 \lesssim 63 \\
            0.059 & 63 \lesssim  \kappa_2^T \lesssim 73 \\
            b_1(\kappa^T_2)^{-\frac{7}{10}} + b_2 & 73\lesssim \kappa^T_2 \lesssim 458 \\
            b_3 \kappa^T_2 +b_4  & 458 \lesssim \kappa^T_2 \lesssim 743,
        \end{cases}
\label{eq:fit_ekfunc}
\end{equation}
with best-fit coefficients $b_1 = 2.44$, $b_2 = -0.019$, $b_3 = -5.1\times10^{-5}$, and $b_4 = 0.038$ (see the technical note at https://dcc.ligo.org/T1800417/public/ for details). Note that the fit is calibrated only up to $\kappa_2^T \simeq 743$, well above the range relevant for GW170817. This fitting formula provides a practical estimate of $E_{\rm GW,p}$, although a fractional uncertainty of $\sim30\%$ should be included.

For such a purpose, the key point is to reliably estimate $\kappa_2^{T}$, which needs to robustly reconstruct the EoSs of the NS matter. In our earlier approach of \citet{Fan:2020hwe}, this task was primarily carried out using parametrized EoS models \citep{Jiang:2019rcw}. Although nonparametric approaches were also explored \citep{Landry:2020vaw}, the tidal deformability was inferred indirectly \citep{Kumar:2019xgp}. In addition, those analyses neither incorporated constraints from chiral effective field theory ($\chi$EFT) at low densities nor perturbative QCD (pQCD) at asymptotically high densities, both of which are now recognized as providing valuable information on the NS EoSs \citep{Gorda:2022jvk,Han:2022rug}. The available observational data at that time were also relatively limited. More recently, \citet{Fan:2023spm} performed a comprehensive Bayesian nonparametric inference of NS EoSs using three independent frameworks: a single-layer feed-forward neural network \citep{Han:2021kjx}, a piecewise-linear parametrization of the sound speed \citep{Jiang:2022tps, Annala:2021gom}, and a Gaussian process (GP) approach \citep{Landry:2018prl, Essick:2019ldf}. The remarkable agreement among the results obtained with these different methods suggests that the inferred EoSs are robust against modeling choices. Motivated by this consistency, we adopt the GP-based inference scheme in this work to evaluate the tidal coupling parameter $\kappa_2^{T}$ relevant for GW170817. Our implementation differs only in the observational input: we employ an updated data set that includes the recent mass–radius measurement of PSR J0437–4715, together with revised constraints for PSRs J0030+0451 and J0740+6620. This update enables a more accurate reassessment of the post-merger GW emission.

Below we briefly summarize the key elements of the GP-based EoS model. We use $\phi$ to represent EoS through the function $\phi(n) \equiv -\ln\!\left(1/c_{\rm s}^2-1\right)$,
where $c_{\rm s}$ is the sound speed and $n$ is the baryon number density. In this framework, $\phi(n)$ is modeled as a GP, which may be viewed as a multivariate Gaussian distribution characterized by a set of hyperparameters. Specifically, we assume
\begin{equation}
\phi(n) \sim \mathcal{N}\!\left(-\ln\!\left(1/\bar{c}_{\rm s}^2-1\right), K(n_i,n_j)\right),
\end{equation}
where the covariance kernel is taken to be $K(n_i,n_j) = \eta^2 \exp\!\left[-\frac{(n_i-n_j)^2}{2l^2}\right]$, with $n_i$ and $n_j$ denoting the baryon number densities of any two points at which the process is evaluated.
The hyperparameters $\{\bar{c}_{\rm s}^2, \eta^2, l\}$ correspond to the mean squared sound speed, the variance, and the correlation length, respectively. We adopt the same prior settings for these parameters as in \citet{Fan:2023spm}. At low densities (up to $\sim 1.1$ times the nuclear saturation density), the EoSs are conditioned to reproduce $\chi$EFT constraints by sampling from the corresponding posterior samples of \citet{Drischler:2017wtt,Drischler:2020hwi}.
At high densities, compatibility with pQCD calculations is enforced through an appropriate likelihood function \citep{Gorda:2022jvk}. Together, these ingredients yield a flexible yet physically grounded reconstruction of the NS EoSs.

\begin{figure}
  \centering
  \includegraphics[width=1.0\textwidth]{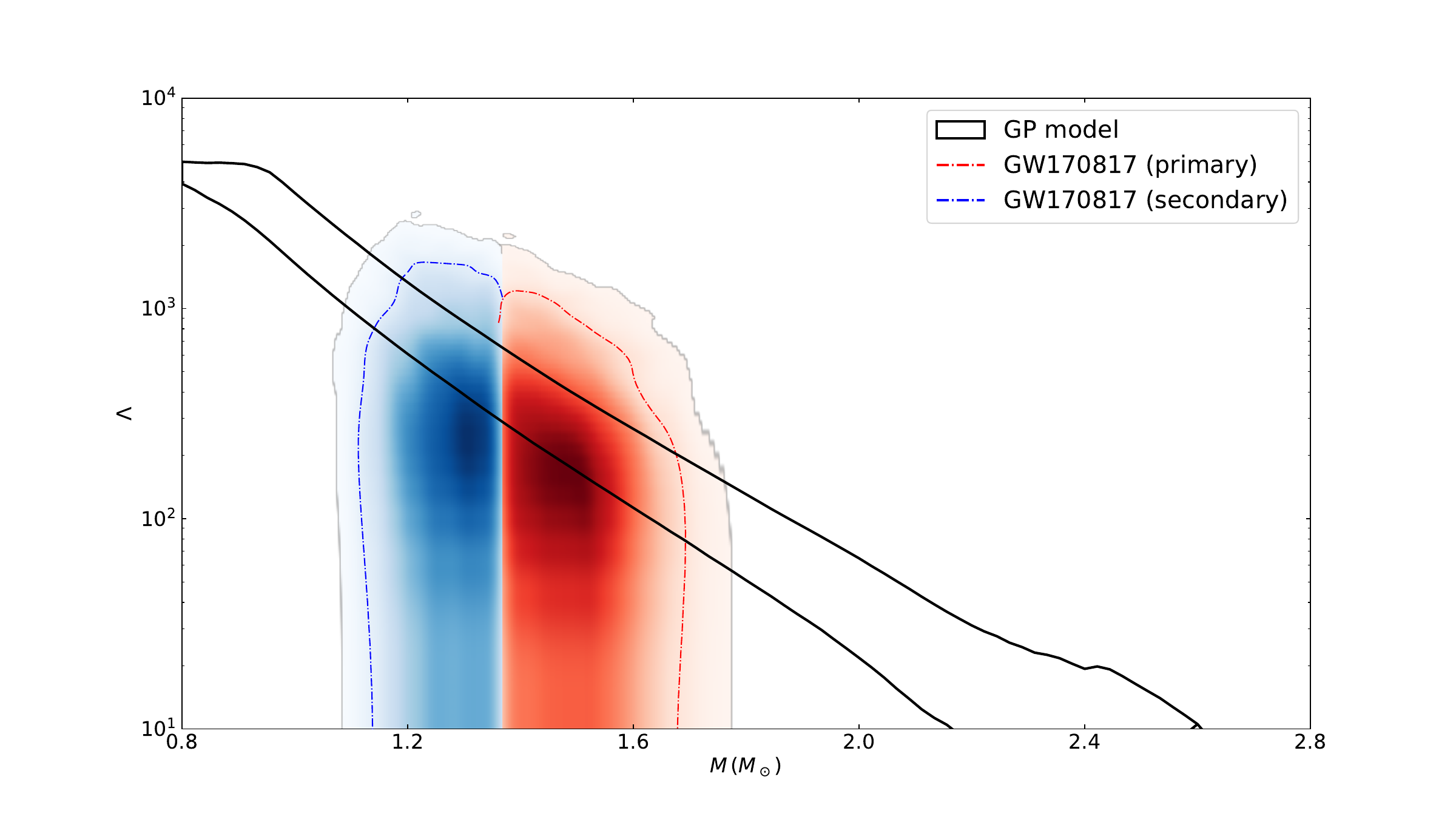}
  \caption{The $90\%$ credible intervals of the mass–tidal deformability relation derived from the reconstructed EoSs, shown in comparison with the direct measurements (dash-dotted lines) on mass and tidal deformability of GW170817 \citep{LIGOScientific:2018mvr}.}
  \label{fig:M_Lambda}
\end{figure}
Fig.~\ref{fig:M_Lambda} presents $90\%$ credible intervals ($\mathrm{CI}$) for the mass–tidal deformability relation inferred from the reconstructed EoSs. These intervals are narrower than those obtained in earlier studies (e.g., \citealt{De:2018uhw, Jiang:2019rcw}). Quantitatively, the tidal deformability of a canonical $1.4\,M_\odot$ NS is now constrained to $\Lambda_{1.4}=408_{-157}^{+165}$ (90\% CI), roughly $50\%$ tighter than the constraints available at the time of \citet{Jiang:2019rcw}. The improvement is mainly driven by the newly included NICER measurement of PSR J0437$-$4715, the updated analyses of PSRs J0740+6620 and J0030+0451, and the incorporation of the $\chi$EFT and pQCD boundary conditions. For GW170817, the component masses of the two NSs are constrained to the ranges $1.36$–$1.60M_\odot$ and $1.16$–$1.36M_\odot$ together with tidal deformability measurements \citep{LIGOScientific:2018hze,LIGOScientific:2018mvr}.
\begin{figure}
  \centering
  \includegraphics[width=1.0\textwidth]{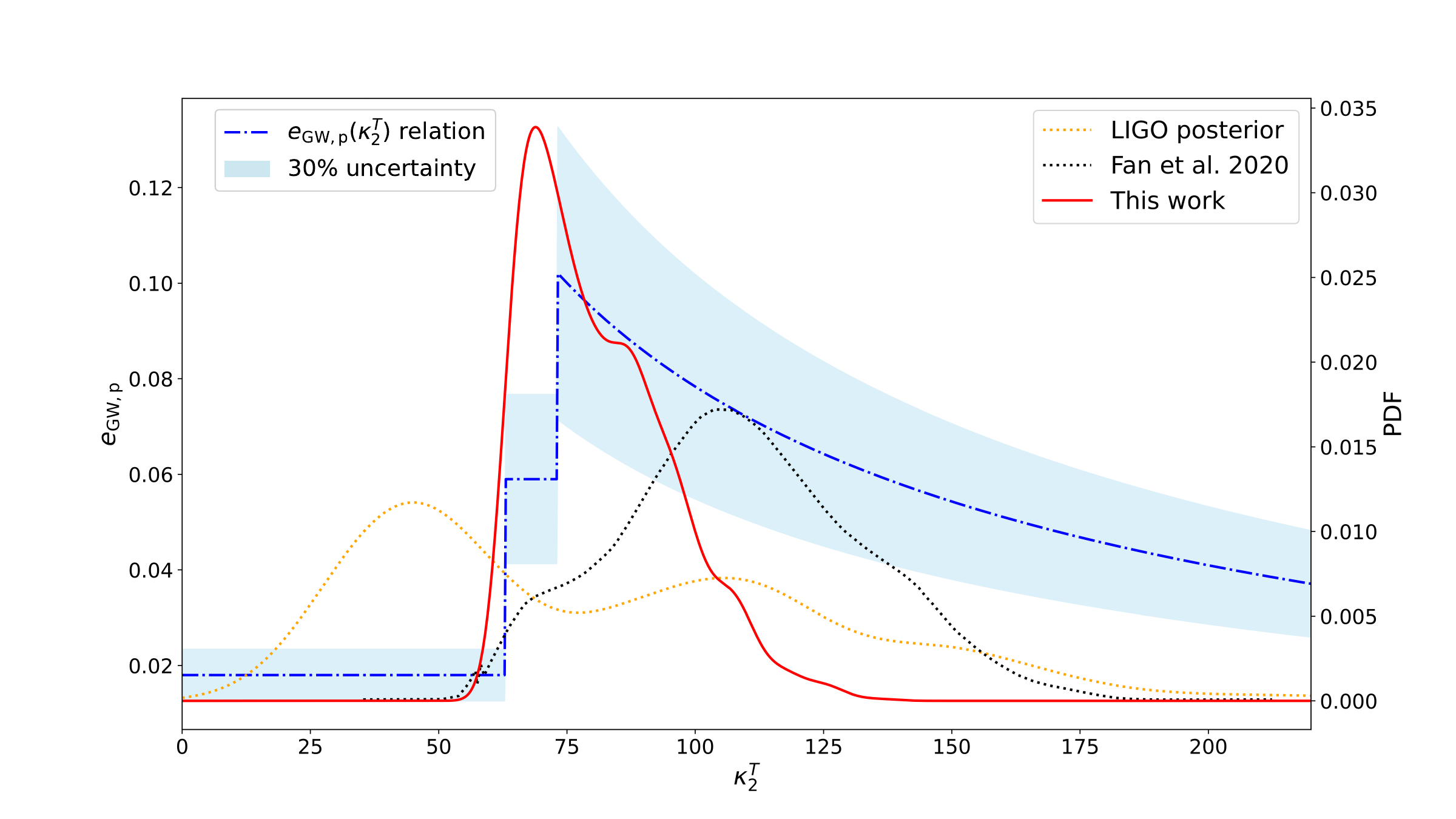}
  \caption{The inferred distribution of $\kappa_2^T$ for GW170817 together with the correlation between $e_{\rm GW,p}$ and $\kappa_2^T$. The dash-dotted blue curve represents the quasi-universal relation given by Eq.~\eqref{eq:fit_ekfunc}, and the shaded blue region denotes the adopted $30\%$ fractional uncertainty. For comparison, the orange dotted curve corresponds to the result directly obtained from the posterior samples of GW170817 \citep{LIGOScientific:2018mvr}, and the black dotted curve shows the result reported by \citet{Fan:2020hwe}.}
  \label{kappa2t}
\end{figure}
Proceeding to infer the $\kappa_2^T$ distribution with these reconstructed EoSs, we adopt a pragmatic approach and employ a representative ensemble of EoSs consistent with current observational constraints. Specifically, we resample $10^4$ EoS samples from more than $10^5$ EoSs, weighted by the constraints from GW observations, NICER measurements, and theoretical bounds from $\chi$EFT and pQCD, following \citet{Tang:2023owf}. Applying these EoSs to NS structure equations, we obtain $10^4$ realizations of the mass–tidal deformability relation $(M,\Lambda)$. Interpolating over the mass posteriors of GW170817 yields the tidal deformabilities of the two NSs and hence the distribution of $\kappa_2^T$. The parameter space with $\kappa_2^T \lesssim 63$ corresponds to scenarios in which the merger remnant collapses to a black hole (BH) within a dynamical timescale after merger, i.e., the so-called prompt collapse. Such prompt-collapse outcomes are incompatible with GW170817, for which multiple lines of evidence indicate a delayed collapse of the remnant \citep{LIGOScientific:2017ync, Metzger:2019zeh, Hajela:2021faz}. We therefore discard posterior samples with $\kappa_2^T < 63$ when analyzing GW170817. After excluding this prompt-collapse regime, we find $\kappa_2^T = 78^{+17}_{-11}$ (68.3\%~\rm CI). As illustrated in Fig.~\ref{kappa2t}, the peak of the $\kappa_2^T$ distribution lies not far from the value that maximizes $e_{\rm GW,p}$, and thus $E_{\rm GW,p}$.
\begin{figure}
  \centering
  \includegraphics[width=1.0\textwidth]{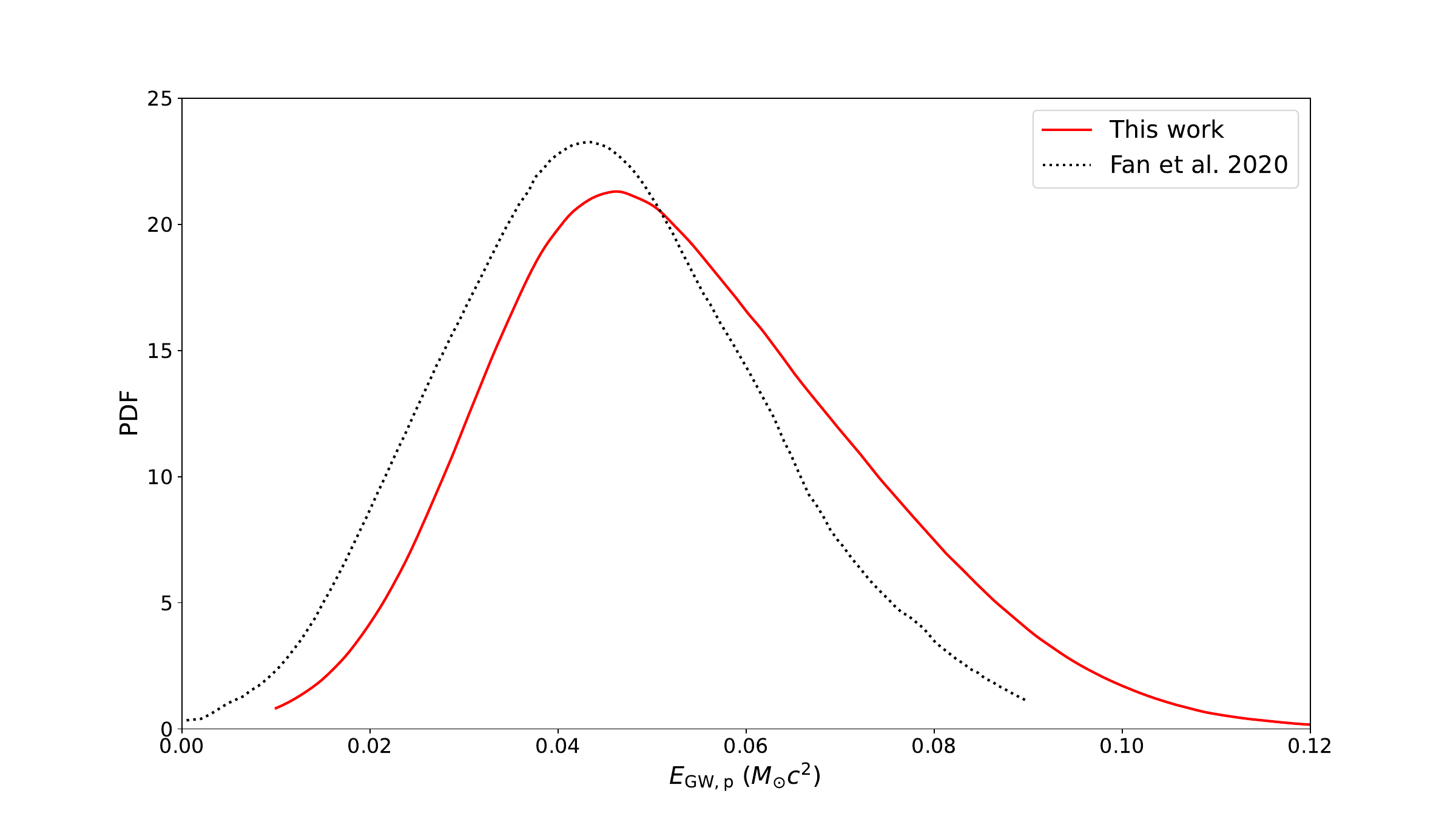}
  \caption{The $E_{\rm GW,p}$ for GW170817 estimated in this work (red line) is $E_{\rm GW,p} = 0.051^{+0.037}_{-0.027}\,M_\odot c^2 ~(90\%~\mathrm{CI})$, which is slightly larger than that found in \citet{Fan:2020hwe}, and confirms that in the post-merger phase GW170817 is indeed an efficient gravitational wave radiator.}
\label{fig:E_gwp}
\end{figure}
This indicates that GW170817 likely belongs to the class of BNS mergers with particularly strong post-merger GW emission. Comparing to the estimate of \citet{Fan:2020hwe}, who obtained $\kappa_2^T = 108^{+41}_{-38}$ at $90\%~\mathrm{CI}$ by the spectral decomposition method \citep{Jiang:2019rcw}, nowadays the inferred distribution is shifted toward lower values, with narrower intervals. Using the posterior distribution of $\kappa_2^T$ for GW170817 and the correlation in Eq.~\eqref{eq:fit_ekfunc}, we estimate the post-merger GW energy. Although $e_{\rm GW,p}$ varies over a relatively broad range with the inferred $\kappa_2^T$, this variation is largely encompassed by the adopted $30\%$ fractional uncertainty. In practice, for each posterior sample we multiply the fitted $e_{\rm GW,p}(\kappa_2^T)$ by a factor $(1+\epsilon)$, where $\epsilon$ is drawn from a zero-mean Gaussian with a standard deviation of $0.3$ (truncated to keep $e_{\rm GW,p}$ positive); this scatter dominates the final error budget of $E_{\rm GW,p}$. Including this uncertainty, we obtain $E_{\rm GW,p} = 0.051^{+0.022}_{-0.017}\,M_\odot c^2$ ($0.051^{+0.037}_{-0.027}\,M_\odot c^2$ at $90\%~\mathrm{CI}$) as illustrated in Fig.~\ref{fig:E_gwp}, corresponding to an energy loss of nearly $2\%$ of the total BNS mass-energy. Despite the significant improvement in the precision of $\kappa_2^T$, the inferred value of $E_{\rm GW,p}$ remains close to previous estimates, as the uncertainty budget is dominated by the intrinsic scatter of the fitting relation.

\section{Constraints on $M_{\rm TOV}$ within the supramassive NS scenario of GW170817}
\label{sec:third}
After the merger of a BNS, the outcome can be classified as follows \citep{Baumgarte:1999cq,Baiotti:2016qnr,Radice:2020ddv,Bernuzzi:2020tgt}: (i) prompt collapse, in which the collapse occurring during the first increasing of the NS central density; (ii) hypermassive NS (HMNS), with a mass exceeding the maximum allowed for uniform rotation and temporarily supported by differential rotation, collapsing within ~10–100 ms; (iii) a supramassive NS (SMNS), supported by rigid rotation, surviving much longer (even seconds) until enough angular momentum is lost; and (iv) a stable NS. The first and last outcomes are not the focus of this work, as multiple lines of evidence suggest a delayed collapse for GW170817; in particular, the collapse time of the remnant has been inferred to be $t_{\rm c}=0.98^{+0.31}_{-0.26}$~s by \citet{Gill:2019bvq}. For a highly magnetized neutron star, the magnetic braking timescale of the differential rotation is just $\sim 0.1$s \citep{Shapiro:2000zh,Gao:2005yd}, so it is reasonable to assume that the remnant of GW170817 was a SMNS rotating uniformly before its collapse if it survived for $\gtrsim0.1$ s. We note, however, that the lifetime of the remnant is still quite uncertain: several recent analyses of the kilonova properties (e.g., the strength of specific spectral features and the inferred electron fraction of the ejecta) favor a considerably shorter lifetime of $\sim0.1$~s or even less \citep{Domoto:2022cqp,Perego:2020evn,Tarumi:2023apl,Sneppen:2024jch,Jacobi:2025eak}, and modern 3D general-relativistic radiation-hydrodynamics simulations reveal additional mass-ejection channels from a long-lived remnant, such as the spiral-wave wind and the neutrino-driven wind \citep{Nedora:2019jhl,Nedora:2020hxc,Just:2023wtj}, that were not included in the prescription of \citet{Gill:2019bvq}. Therefore, in addition to the fiducial case anchored to the \citet{Gill:2019bvq} estimate, we also treat $t_{\rm c}$ as a free parameter and explore the limiting cases $t_{\rm c}=0.1$, $0.3$, and $1.0$~s. Therefore one can estimate $M_{\rm TOV}$ with GW170817, as long as the angular momentum and mass loss of the remnant can be reliably evaluated. For such a purpose, we mainly follow the approaches of \citet{Shibata:2019ctb} and \citet{Shao:2019ioq}.

Assuming conservation of baryon mass from the inspiral phase up to the moment of collapse, we impose the following relation, $M_{\rm b,tot} = M_{\rm b,crit} + m_\mathrm{loss}$, where the subscript $b$ denotes baryonic (rest) mass, and `crit' refers to the collapse moment, when we identify a SMNS at the turning point along an equilibrium sequence of rotating NS configurations \citep{Shibata:2019ctb,Shao:2019ioq}. The term $m_\mathrm{loss}$ accounts for all the baryonic material outside the SMNS at the collapse moment,
\begin{equation}
m_{\rm loss} = M_{\rm eje} + M_{\rm torus}, \qquad
M_{\rm eje} = M_{\rm dyn} + M_{\nu\rm -wind}^{\rm NS} + M_{B\rm -wind}^{\rm NS} + M_{\nu\rm -wind}^{\rm disk} + M_{B\rm -wind}^{\rm disk},
\label{eq:mloss}
\end{equation}
where $M_{\rm dyn}$ is the dynamical ejecta launched within a dynamical timescale, $M_{\nu(B)\rm -wind}^{\rm NS}$ are the neutrino-driven (magnetically driven) winds from the remnant NS, and $M_{\nu(B)\rm -wind}^{\rm disk}$ are the corresponding winds from the disk. Adopting the mass-ejection prescription of \citet{Gill:2019bvq}, these components can be parameterized by $t_\mathrm{c}$. Following \citet{Wang:2018nye}, the corresponding torus mass is inferred to lie in the range $M_{\rm torus} \simeq 0.015$–$0.134\,M_\odot$ at the $90\%$ confidence level, with a most probable value of $M_{\rm torus} \simeq 0.035\,M_\odot$. Strictly speaking, the constraint of \citet{Wang:2018nye} applies to the torus mass right after BH formation, since it is derived from the energetics of the gamma-ray burst jet and the kilonova. For the delayed-collapse scenario considered here ($t_{\rm c}\sim0.1$--$1$ s), however, this is a reasonable proxy for the torus mass at the collapse moment: by that time the torus has viscously spread to $R_{\rm torus}\sim70$--$140$ km, far outside the innermost stable circular orbit ($\sim12$--$16$ km) of the newly formed BH, so only the innermost material smoothly connected to the stellar surface is promptly swallowed at BH formation. To quantify the impact of a possible prompt fall-back, we introduce a swallowed fraction $f_{\rm sw}$ such that $M_{\rm torus}({\rm collapse}) = M_{\rm torus}({\rm observed})/(1-f_{\rm sw})$, and find that even the extreme choice $f_{\rm sw}=0.5$ only shifts the inferred $M_{\rm TOV}$ by $\Delta M_{\rm TOV}\simeq+0.03\,M_\odot$, well within the statistical uncertainty. Although baryon mass conservation is conceptually straightforward, a nontrivial issue arises when converting the observed gravitational masses of NSs into baryonic masses, as this conversion depends on the NS binding energy. Denoting the gravitational masses of the two progenitor NSs by $M_1$ and $M_2$, the total baryonic mass can be written as $M_{\rm b,tot} = M_1 + M_2 + {\rm BE}_1 + {\rm BE}_2$, where ${\rm BE}$ is the (positive) binding energy of a NS. For nonrotating (or slowly rotating) NSs with compactness in the range $0.05 \le \mathscr{C} \le 0.25$, the binding energy satisfies the empirical relation ${\rm BE}/M = -0.0130 + 0.618\mathscr{C} + 0.267\mathscr{C}^2$ \citep{Shao:2019ioq}. At the moment of collapse, the remnant is assumed to be a SMNS located at the turning point of an equilibrium sequence of uniformly rotating configurations, such that $M_{\rm b,crit} = M_\mathrm{crit} + {\rm BE}_\mathrm{crit}$. For a rapidly rotating NS at the turning point, the corresponding binding energy relation becomes ${\rm BE}_\mathrm{crit}/M_\mathrm{crit}= -0.10 + 0.78(1 - 0.050j - 0.034j^2)\mathscr{C}_{\mathrm{TOV}} + 0.61(1 + 0.23j - 0.58j^2)\mathscr{C}_{\mathrm{TOV}}^2$ \citep{Shao:2019ioq}, where $j \equiv cJ/(GM^2)$ is the dimensionless angular momentum, $J$ is the angular momentum of the remnant, and $\mathscr{C}_\mathrm{TOV} = GM_\mathrm{TOV}/(R_\mathrm{TOV}c^2)$ denotes the compactness of the maximum-mass nonrotating configuration. Importantly, \citet{Shao:2019ioq} found a QUR between $M_\mathrm{crit}/M_{\mathrm{TOV}}$ and $j$, which is $M_\mathrm{crit}/M_\mathrm{TOV} = 1 + 9.02\times10^{-2}\mathscr{C}_\mathrm{TOV}^{-1}j^2+1.93\times10^{-2} \mathscr{C}_\mathrm{TOV}^{-2} j^4$ (see also \citet{Breu:2016ufb} for a similar expression).

To robustly estimate $M_{\rm TOV}$, we need a reasonable evaluation of the angular momentum at the collapse time, which reads $J_{\rm coll} = J_0 - J_{\rm eje} - J_{\rm torus} - J_{\rm GW,p} - J_{\nu}$, a variant of the global angular momentum balance equation. We stress that this equation expresses the angular momentum conservation of the whole system rather than a torque balance on the central remnant; since the torus term is anchored to the post-wind torus mass, the angular momentum carried away by the disk winds must be counted separately in $J_{\rm eje}$, otherwise the initial angular momentum of the disk would be underestimated. $J_0$ is the total angular momentum of the binary NS system at the onset of merger (i.e., at the moment when the two stars come into contact) and can be estimated as $J_0 \approx Gc^{-1} M_{\rm tot}^2 \nu \left[a_1 - a_2 \delta\nu + a_3 \left(\frac{R_{1.35}}{10{\rm km}}\right)^3 (1 + a_4 \delta\nu)\right]$, where $\delta\nu=\nu-1/4$, $a_1\simeq3.32$, $a_2\simeq31$, $a_3\simeq0.137$, and $a_4\simeq27$ are the fitting coefficients, and $R_{1.35}$ denotes the radius of a $1.35M_\odot$ NS \citep{Shibata:2019ctb}. $J_{\rm eje}$ denotes the angular momentum carried away by mass ejection and can be approximated as $J_{\rm eje} \simeq M_{\rm eje}\sqrt{G M_{\rm crit} R_{\rm eje}}$, where $R_{\rm eje}$ is the characteristic radius at which the ejecta are launched. For the dynamical ejecta launched on a dynamical timescale, $R_{\rm eje}$ can be estimated as the typical radius of the surrounding torus at early times, which we set to $50~\mathrm{km}$ in the case. We assign $R_{\rm eje} = R_{\rm crit}$ for ejecta launched directly from the remnant. In addition, both neutrino-driven and magnetically driven ejecta predominantly originate from the torus at later times, for which we take $R_{\rm eje}=R_{\rm torus}$, where $R_{\rm torus} \approx 40+100(t_{\rm c}/{\rm 1s})^{1/2}~{\rm km}$ \citep{Fujibayashi:2017puw} denotes the characteristic torus radius at the moment of collapse. Consistently, the angular momentum of the torus itself is estimated as $J_{\rm torus} \simeq M_{\rm torus}\sqrt{G M_{\rm crit} R_{\rm torus}}$ (see also  the disk mass--angular momentum relation recently obtained from numerical-relativity simulations \citep{Camilletti:2024otr}). The angular momentum radiated away via post-merger GWs is $J_{\rm GW,p} \approx E_{\rm GW,p}/\pi f \approx 9.5 \times 10^{48} {\rm erg\,s}~(E_{\rm GW,p}/0.05M_\odot c^2)(f/3{\rm kHz})^{-1}$, with $f$ as the dominant frequency of GWs during the post-merger phase. The neutrino-carried angular momentum is approximated as $J_{\nu} \simeq 2.2\times10^{48} {\rm erg s} \left(\frac{E_{\nu}}{0.1 M_\odot c^2}\right) \left(\frac{R_{\rm crit}}{13 {\rm km}}\right)^2 \left(\frac{\Omega}{10^4 {\rm rad s}^{-1}}\right)$ \citep{Baumgarte:1998sn}, where $E_\nu=L_\nu t_{\rm c}\sim 2\times 10^{53}\,{\rm erg\,s^{-1}}\,t_{\rm c}$ is the energy of neutrino radiation \citep{Foucart:2015gaa,Radice:2016dwd,Sekiguchi:2016bjd} and $\Omega$ is the angular velocity of the remnant. Finally, analogous to the QUR for $M_{\rm crit}/M_{\rm TOV}$, the QUR connecting $R_{\rm crit}$ and angular momentum can be written as $R_{\rm crit}/R_{\rm TOV} = 1 + 0.032\mathscr{C}_{\rm TOV}^{-1.6} j^2 + 0.014\mathscr{C}_{\rm TOV}^{-3.2} j^4$, where $j$ should be evaluated at the collapse moment, i.e., $j=j_{\rm coll}=cJ_{\rm coll}/(GM_{\rm crit}^2)$.

\begin{figure}
  \centering
  \includegraphics[width=0.7\textwidth]{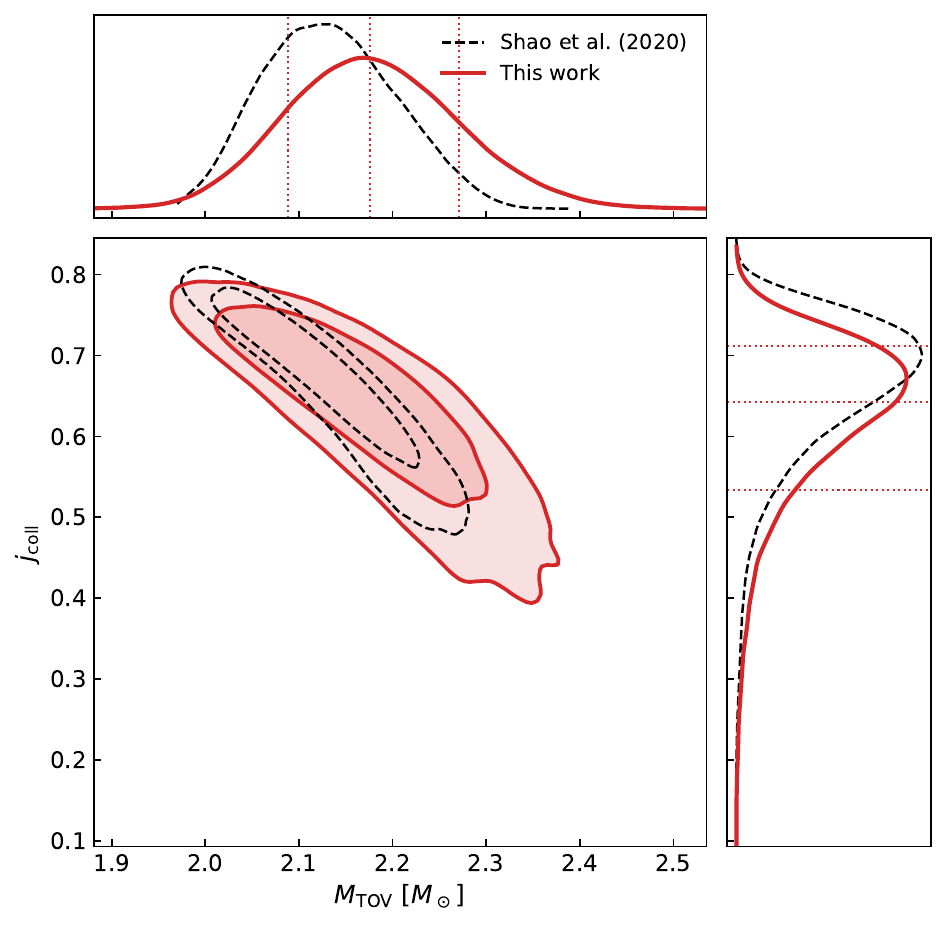}
  \caption{Joint posterior distribution of $M_{\rm TOV}$ and $j_{\rm coll}$ inferred from GW170817 (fiducial case). The central panel shows the two-dimensional joint distribution, enclosed by the contours of $68.3\%$ and $90\%~\mathrm{CI}$, while the upper and right panels display the corresponding marginalized distributions. The results obtained by \citet{Shao:2019ioq} are also shown for comparison and our current $M_{\rm TOV}$ shifts toward a higher value.}
  \label{fig:Mtov_j}
\end{figure}
With the above ingredients, we are able to extract the information on the maximum mass of a nonrotating NS from its rotating counterpart at the critical state. For convenience, we denote the binding energy of a progenitor NS as ${\rm BE}=f(M,\mathscr{C})$, and the binding energy at collapse as ${\rm BE}_{\rm crit}=g(M_{\rm crit},\mathscr{C}_{\rm TOV},j_{\rm coll})$. Substituting the definitions of $M_{\rm b,tot}$ and $M_{\rm b, crit}$ into baryon mass conservation relation, we obtain
\begin{equation}
\label{eq:f}
g(M_{\rm crit},\mathscr{C}_{\rm TOV},j_{\rm coll}) + M_{\rm crit}
= M_{\rm tot} - m_{\rm loss} + f(M_1,\mathscr{C}_1) + f(M_2,\mathscr{C}_2),
\end{equation}
where the right-hand side depends on the specific merger scenario and must be evaluated on a case-by-case basis. By jointly solving this equation together with the global angular momentum balance equation, the pair $(M_{\rm TOV},j_{\rm coll})$ can be inferred self-consistently (see also \citealp{Shao:2019ioq} for an approximately analytic treatment). In Sec.~\ref{sec:second}, we obtained an improved estimate of the post-merger GW energy $E_{\rm GW,p}$ for GW170817 by exploiting recently reconstructed NS EoSs and the corresponding QURs. It is therefore worth reacquiring the joint distribution of ($M_{\rm TOV}$, $j_{\rm coll}$). The results are shown in Fig.~\ref{fig:Mtov_j}. The marginalized distribution of $M_{\rm TOV}$ is constrained to $2.18\pm0.09\,M_{\odot}$ ($2.18^{+0.16}_{-0.14}\,M_{\odot}$ at $90\%~\mathrm{CI}$), which is consistent with the latest result reported by \citet{Tang:2026yta}. Meanwhile, the marginalized distribution of the dimensionless angular momentum at collapse is $j_{\rm coll} = 0.64^{+0.07}_{-0.11}$.

\begin{figure}
  \centering
  \includegraphics[width=1.0\textwidth]{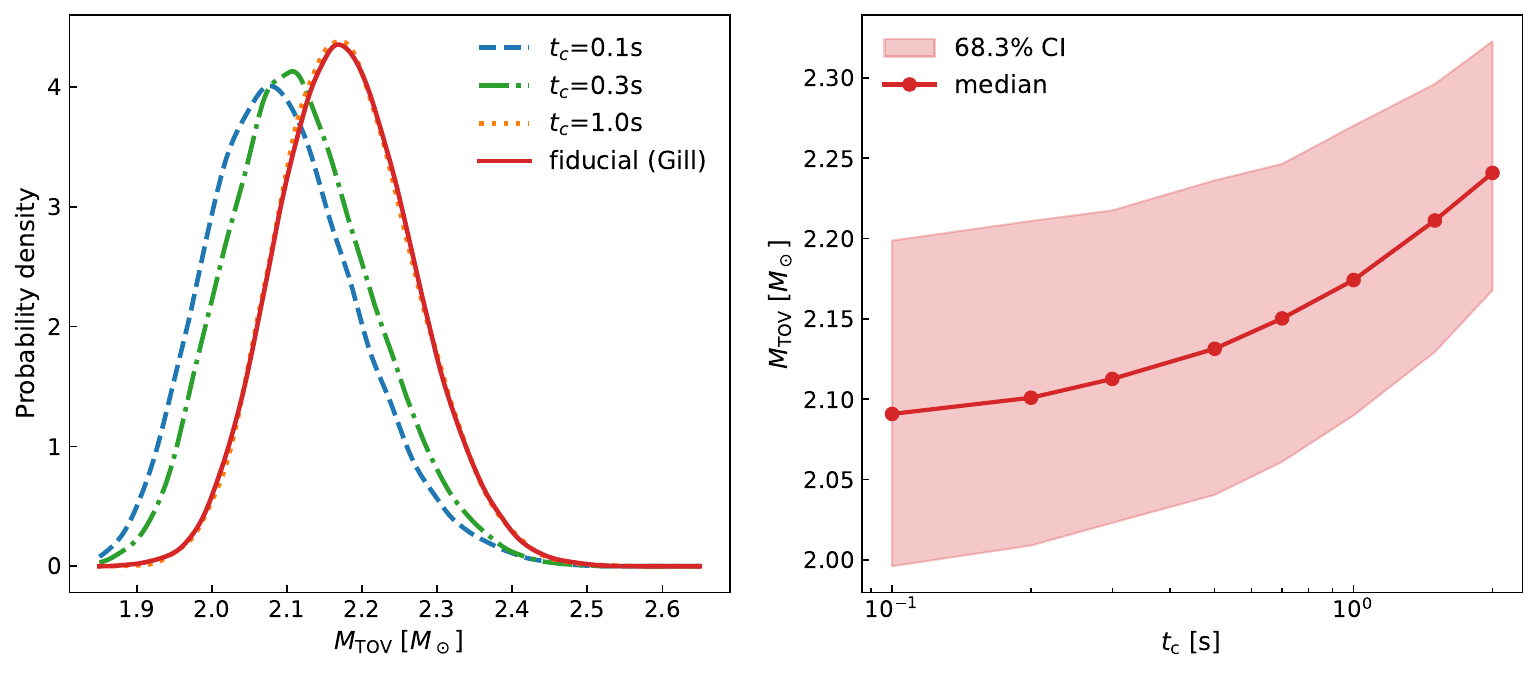}
  \caption{Dependence of the inferred $M_{\rm TOV}$ on the assumed collapse time $t_{\rm c}$. Left: marginalized posterior distributions of $M_{\rm TOV}$ for $t_{\rm c}=0.1$, $0.3$, and $1.0$~s, together with the fiducial case based on the estimate of \citet{Gill:2019bvq}. Right: median (dots) and $68.3\%$ credible interval (shaded band) of $M_{\rm TOV}$ as a function of $t_{\rm c}$.}
  \label{fig:tc_dep}
\end{figure}

\begin{figure}
  \centering
  \includegraphics[width=0.75\textwidth]{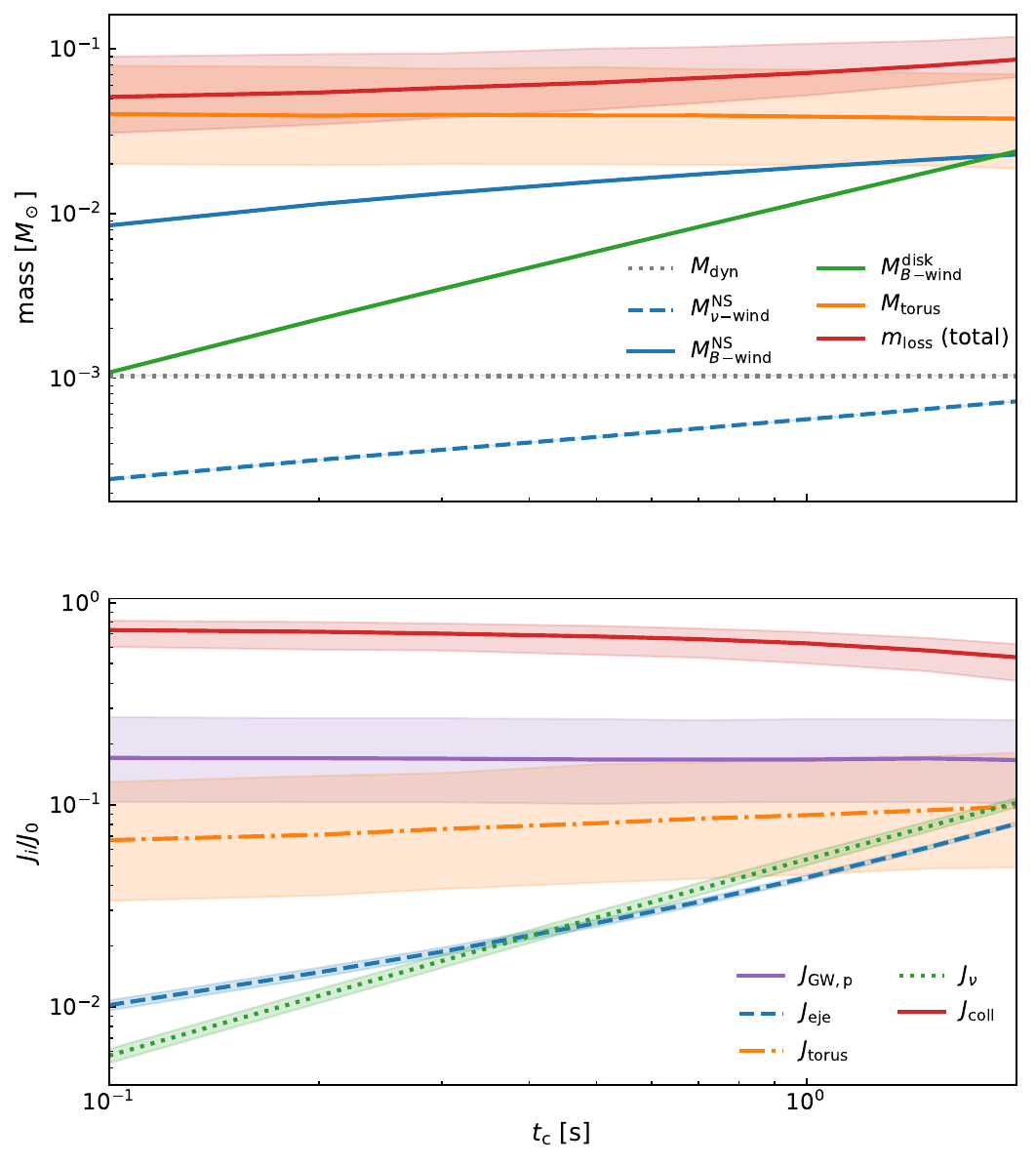}
  \caption{The mass (top) and angular momentum (bottom, normalized to $J_0$) budgets as functions of the collapse time $t_{\rm c}$. Solid/dashed lines mark the medians and the shaded bands the $68.3\%$ credible intervals.}
  \label{fig:budget}
\end{figure}

Given the uncertainty of the remnant lifetime discussed above, we further explore the dependence of the results on $t_{\rm c}$, treated as a free parameter in the range of $0.1$--$2$ s. The results are presented in Fig.~\ref{fig:tc_dep}, and the corresponding mass and angular momentum budgets are shown in Fig.~\ref{fig:budget}. A shorter lifetime implies less mass and angular momentum losses (dominated by the reduced magnetically driven winds and neutrino emission), hence a larger $j_{\rm coll}$ and, for a fixed baryonic mass, a smaller $M_{\rm TOV}$. Quantitatively, we find $M_{\rm TOV}=2.09^{+0.11}_{-0.09}\,M_\odot$, $2.11^{+0.11}_{-0.09}\,M_\odot$, and $2.18^{+0.10}_{-0.09}\,M_\odot$ for $t_{\rm c}=0.1$, $0.3$, and $1.0$~s, respectively, i.e., the systematic uncertainty introduced by the poorly known collapse time ($\sim0.09\,M_\odot$ between the limiting cases) is comparable to the statistical one. If the remnant survived for as long as $2$~s, the inferred $M_{\rm TOV}$ would increase to $2.24^{+0.08}_{-0.07}\,M_\odot$. These limiting cases bracket the plausible range of $M_{\rm TOV}$ given the current uncertainty of the ejection/lifetime modeling.

\begin{figure}
  \centering
  \includegraphics[width=1.0\textwidth]{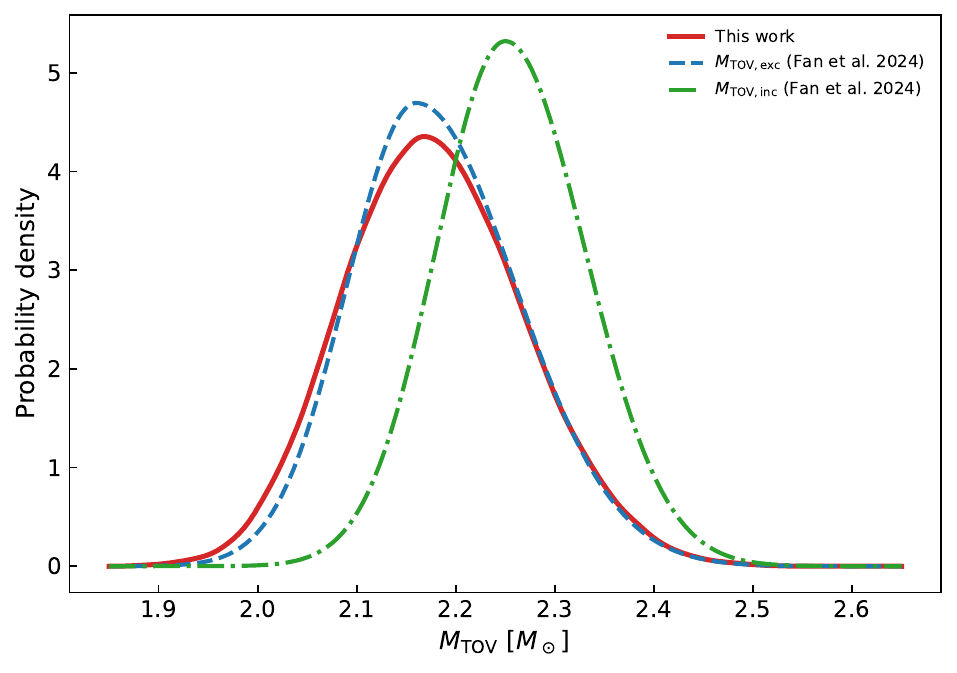}
  \caption{Comparison of the currently inferred $M_{\rm TOV}$ with that found by \citet{Fan:2023spm} with and without incorporating some very massive neutron stars with the masses indirectly measured (i.e., the $M_{\rm TOV,inc}$ and $M_{\rm TOV,exc}$). Our current inference of $M_{\rm TOV}$ is more consistent with $M_{\rm TOV,exc}$ rather than $M_{\rm TOV,inc}$, which calls for the double check of these indirectly measured masses of the neutron stars.
  }
  \label{fig:Mtov_sumary}
\end{figure}

In this section we have found $M_{\rm TOV}=2.18\pm0.09M_\odot$ with the data of GW170817 under the assumption of the collapse of the SMNS remnant at $\sim 1$s after the merger of the BNS. The $M_{\rm TOV}$ of NSs can also be robustly inferred from either the mass distribution of these objects or the reconstruction of the EoSs of the very dense matter. Benefiting from a joint approach of these two methods, \citet{Fan:2023spm} have yielded a precise inference of $M_{\rm TOV,inc} = 2.25^{+0.08}_{-0.07}M_\odot$ (including some very massive NSs measured indirectly) or $M_{\rm TOV,exc} = 2.16^{+0.10}_{-0.07}M_\odot$ (excluding the NSs with masses measured indirectly), all these values are at the $68.3\%~\mathrm{CI}$. We compare the posterior distributions of these three values in Fig.~\ref{fig:Mtov_sumary}. Evidently, the $M_{\rm TOV}$ inferred in this work is well consistent with $M_{\rm TOV,exc}$, but is slightly below $M_{\rm TOV,inc}$. There are three possibilities. One is that the masses of the very massive NSs reported in the literature via the indirect measurements have been overestimated. The second is that the nascent NSs are so hot that the empirical relations developed in the literature, as adopted in this work, are not accurate. The third is that the actual collapse time of the SMNS remnant is longer than the assumed $\sim 1$~s. At present, all three possibilities remain open, though the third is less favored. If NSs as massive as $2.2M_\odot$ can be precisely/directly measured in the future, the second possibility will be favored. If so, the comparison of the masses of the non-rotating NSs inferred in independent ways may shed valuable light on the physical processes taking place inside the nascent NSs.

\section{Summary}
In this work, we have updated the constraint on the expected post-merger gravitational radiation energy of GW170817 and hence the maximum mass of the non-rotating neutron stars assuming the nascent remnant was a supramassive neutron star that collapsed at $t\sim 1$s after the merger. A key improvement of this study lies in the updated estimation of the post-merger GW radiation of GW170817. By employing recently reconstructed NS EoSs, constrained jointly by GW observations, NICER measurements, and theoretical bounds from $\chi$EFT and pQCD, we obtained a refined posterior distribution of the EOB tidal parameter $78^{+17}_{-11}$ for GW170817. Using the calibrated correlation between $\kappa_2^T$ and the dimensionless post-merger GW energy \citep{Zappa:2017xba}, we inferred $E_{\rm GW,p} \simeq 0.051^{+0.037}_{-0.027}\,M_\odot c^2$, which is slightly larger than that found in \citet{Fan:2020hwe} and corresponds to an energy loss of nearly $2\%$ of the total BNS mass-energy. This result indicates that GW170817 belongs to the class of mergers with particularly efficient post-merger GW emission, in favour of a relatively high $M_{\rm TOV}$. Incorporating this updated $E_{\rm GW,p}$ into the angular momentum budget of the merger remnant, and adopting the same prescriptions for mass loss, neutrino emission, ejecta, and torus properties as in \citet{Shao:2019ioq}, we derived the joint posterior distribution of $(M_{\rm TOV}, j_{\rm coll})$. The resulting marginalized constraint, $M_{\rm TOV} = 2.18\pm0.09\,M_{\odot}$, is consistent with that independently inferred from the re-construction of the equation of state of neutron star matter, i.e., $M_{\rm TOV,exc}=2.16^{+0.10}_{-0.07}M_\odot$, particularly if the very massive neutron stars with masses measured indirectly have been removed in constructing the prior distribution of $M_{\rm TOV}$ (if included, the inferred value is enhanced to $M_{\rm TOV,inc}=2.25^{+0.08}_{-0.07}M_\odot$), as found in \citet{Fan:2023spm}. Since the lifetime of the remnant of GW170817 is still uncertain, we have also treated the collapse time as a free parameter and found that the inferred $M_{\rm TOV}$ varies from $2.09^{+0.11}_{-0.09}\,M_\odot$ ($t_{\rm c}=0.1$ s) to $2.18^{+0.10}_{-0.09}\,M_\odot$ ($t_{\rm c}=1$ s); a robust identification of the collapse time of the remnant is thus crucial for a more accurate determination of $M_{\rm TOV}$ in this approach. The consistency of the maximum mass of nonrotating neutron stars found in different approaches suggests a reasonable understanding of this key parameter for dense-matter physics.

Several sources of uncertainty remain in our analysis. These include the intrinsic scatter of the QURs, the modeling of post-merger GW emission (particularly in the transition region of $\kappa_2^T$), the lifetime of the remnant, and assumptions regarding ejecta and torus properties. Nevertheless, our results demonstrate that even in the absence of a direct detection of post-merger GW signals, valuable constraints on $M_{\rm TOV}$ can be obtained by combining inspiral information with numerically calibrated post-merger relations. Looking ahead, this framework can be naturally extended to future BNS events with improved post-merger sensitivity, as expected from next-generation GW detectors \citep{Punturo:2010zz,LIGOScientific:2016wof}. A simultaneous measurement of $\kappa_2^T$, $E_{\rm GW,p}$, and electromagnetic counterparts would significantly reduce current uncertainties and allow for a more direct inference of $(M_{\rm crit}, M_{\rm TOV})$. Moreover, the application of similar methods to systems at the boundary between NSs and BHs, such as GW190814, may further illuminate the nature of the most massive compact objects \citep{Nathanail:2021tay} and the efficiency of GW emission in NS mergers. Overall, our study highlights the crucial role of post-merger physics in NS astrophysics and demonstrates that combining multimessenger observations with QURs provides a powerful avenue for probing the fundamental properties of dense matter.

\begin{acknowledgments}
This work is supported by the National Natural Science Foundation of China under Grants No. 12233011 and No. 12303056, the Project for Young Scientists in Basic Research (No. YSBR-088) of the Chinese Academy of Sciences, and the Postdoctoral Fellowship Program of China Postdoctoral Science Foundation (GZC20241915).
\end{acknowledgments}

\bibliography{ref.bib}{}
\bibliographystyle{aasjournal}

\end{document}